\documentclass[journal]{IEEEtran}

\ifCLASSINFOpdf
\else
   \usepackage[dvips]{graphicx}
\fi
\usepackage{url}

\usepackage{graphicx}
\usepackage{amsmath,times,booktabs,tabularx}
\usepackage[bookmarks=false, hidelinks]{hyperref}
\usepackage{relsize}
\usepackage{color}
\usepackage{amssymb}
\usepackage{array}
\usepackage{textcomp}
\usepackage{multirow}
\usepackage{microtype}
\usepackage{dsfont}
\usepackage[pagewise,switch]{lineno}
\usepackage{mathtools}
\usepackage[space]{cite}

\let\norm\undefined 
\DeclarePairedDelimiter\norm{\lVert}{\rVert}

\DeclareMathOperator*{\argmax}{arg\,max}
\definecolor{mygray}{gray}{0.5}
\newcommand{\rpm}{\raisebox{.2ex}{$\scriptstyle\pm$}}

\begin{document}

\title{Improving Sound Event Classification by Increasing Shift Invariance in Convolutional Neural Networks}

\author{Eduardo Fonseca, \IEEEmembership{Student Member, IEEE}, Andres Ferraro, Xavier Serra
\thanks{Authors are with the Music Technology Group at Universitat Pompeu Fabra, Barcelona, Spain. E-mail for all authors: (name.surname@upf.edu)}
}

\maketitle

\begin{abstract}
Recent studies have put into question the commonly assumed shift invariance property of convolutional networks, showing that small shifts in the input can affect the output predictions substantially.
In this paper, we analyze the benefits of addressing lack of shift invariance in CNN-based sound event classification.
Specifically, we evaluate two pooling methods to improve shift invariance in CNNs, based on low-pass filtering and adaptive sampling of incoming feature maps.
These methods are implemented via small architectural modifications inserted into the pooling layers of CNNs.
We evaluate the effect of these architectural changes on the FSD50K dataset using models of different capacity and in presence of strong regularization.
We show that these modifications consistently improve sound event classification in all cases considered.
We also demonstrate empirically that the proposed pooling methods increase shift invariance in the network, making it more robust against time/frequency shifts in input spectrograms.
This is achieved by adding a negligible amount of trainable parameters, which makes these methods an appealing alternative to conventional pooling layers.
The outcome is a new state-of-the-art mAP of 0.541 on the FSD50K classification benchmark.
\end{abstract}
\begin{IEEEkeywords}
Shift invariance, sound event classification, low-pass filtering, adaptive polyphase sampling, convolutional neural networks
\end{IEEEkeywords}

\IEEEpeerreviewmaketitle

\section{Introduction}
\label{sec:intro}
\IEEEPARstart{C}{onvolutional} Neural Networks (CNNs) have been one of the cornerstones of Sound Event Classification or Tagging (SET) in recent years \cite{fonseca2020addressing,kong2019panns,gong2021psla,Fonseca2019learning}.
One of their commonly assumed properties is \textit{shift} or \textit{translation invariance}, by which output predictions are not affected by small shifts (or even small deformations) in the input signal. 
In theory, this is ensured by the convolution and pooling operations forming the CNNs.
However, recent works in computer vision uncover that this is not always the case.
Azulay and Weiss find that small shifts and transformations in the input can change the network’s predictions substantially \cite{azulay2018deep}.
In particular, they quantify that by shifting or resizing a random input image by one single pixel
, the top class predicted can change with a probability of up to 15\% and 30\%, respectively.
This and other related works \cite{engstrom2018rotation,zhang2019making} empirically show the brittleness of CNNs against minor input perturbations, and their only-partial invariance to shifts.

These works argue that one of the causes of the lack of shift invariance is a wrongly executed subsampling operation that ignores the classic sampling theorem. 
This theorem establishes that, for the subsampling to be done correctly, the sampling rate must be at least twice the highest frequency in the incoming signal \cite{oppenheim2001discrete}.
Otherwise, \textit{aliasing} problems can occur, generating lack of shift invariance in the system and potentially causing a certain distortion in the output---some of the highest frequency components can overlay other low frequency ones.
To address this issue, the classical signal processing measure is to introduce an \textit{anti-aliasing} low-pass filter before downsampling in order to limit the signal’s band \cite{oppenheim2001discrete}. 
In CNNs, subsampling operations are prevalent through strided layers, e.g., convolution or pooling layers with a stride larger than one.
As anti-aliasing actions are not usually taken, feature maps containing high frequency components may lead to shift invariance and/or distortion problems.

The findings above have led to a growing area of research aimed at increasing shift invariance in CNNs, either through architectural improvements \cite{zhang2019making,chaman2021truly,vasconcelos2020effective} or via data augmentation \cite{engstrom2018rotation}. In this work, we are interested in the former, which usually revolves around the idea of improving the subsampling operations.
The predominant trend consists of adding anti-aliasing measures to the CNN architectures.
Similarly to the signal processing fix, some works adopt different low-pass filter based solutions, mainly for image recognition \cite{zhang2019making,vasconcelos2020effective} and more recently also for speech recognition \cite{bruguier2020anti}.
Zhang demonstrates that adding blurring to deep convolutional networks before the strided operations (convolution and pooling) provides increased accuracy on ImageNet \cite{deng2009imagenet} and improved robustness to image perturbations \cite{zhang2019making}.
Vasconcelos et al. conduct a study to isolate the impact of aliasing within the different modules of a ResNet-50 architecture \cite{vasconcelos2020effective}. 
Bruguier et al. insert 1D low-pass filters along the temporal dimension of feature maps in a RNN-based speech recognizer \cite{bruguier2020anti}.

In contrast to the anti-aliasing line of work, another alternative is to design architectural changes to explicitly enforce invariance in the network.
For example, several previous works focus on increasing the invariance of CNNs to \textit{rotations} in input images, by applying constraints to the convolutional filters \cite{worrall2017harmonic} or proposing \textit{ad hoc} operations to enforce this property \cite{dieleman2016exploiting}. 
Recently, to address the lack of shift invariance caused by subsampling operations, Chaman and Dokmanic propose a downsampling mechanism called \textit{adaptive polyphase sampling} \cite{chaman2021truly}.
The key idea is to avoid using the same fixed sampling grid for subsampling a feature map (as typically done in CNNs), but instead select it adaptively based on some criterion (e.g., choosing the grid that produces a downsampled output with highest energy).
To our knowledge, this kind of techniques aimed at fostering shift invariance in CNNs have not been evaluated for sound event classification.

In this paper, we ask whether lack of shift invariance is a problem in sound event recognition, and whether there are benefits in addressing it.
To this end, we apply several mechanisms aimed at increasing shift invariance in the subsampling operations of CNNs, and evaluate them on a large-vocabulary sound event classification task.
Specifically, we adopt mechanisms from the two trends mentioned above, namely, low-pass filters (non-trainable as proposed in \cite{zhang2019making}, as well as a trainable version proposed by us), and adaptive polyphase sampling \cite{chaman2021truly}.
We insert these architectural changes into the max-pooling layers of VGG variants \cite{simonyan2014very}, and we evaluate their effect on the FSD50K dataset \cite{fonseca2020fsd50k} using models of small and large capacity, and in presence of a strong regularizer (\textit{mixup} augmentation \cite{zhang2017mixup}).
We show that these simple changes consistently improve sound event classification in all cases considered.
We also demonstrate that they increase network's robustness to spectrogram shifts.
This is achieved without adding any (or adding very few) trainable parameters, which makes the proposed pooling mechanisms an appealing alternative to conventional pooling layers.
The outcome is a new state-of-the-art mAP of 0.541 on the FSD50K classification benchmark when not using external training data.
Code will be made available in the final version of the paper.\footnote{\url{https://github.com/edufonseca/shift_sec}}

\section{Method}
\label{sec:method}
\begin{figure}[t]
  \centering
  \centerline{\includegraphics[width=0.82\columnwidth]{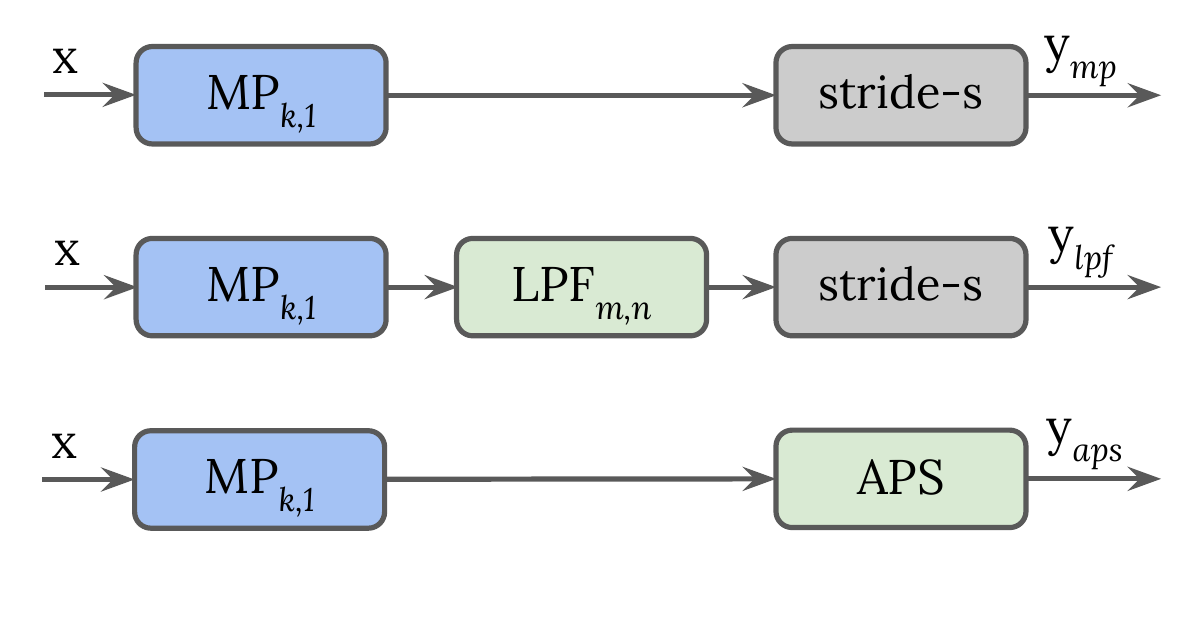}}
  \vspace{-3mm}
  \caption{Max-pooling layer and proposed methods to improve shift invariance. \textit{Top}: A max-pooling layer can be decomposed into a densely-evaluated max-pooling operation with size $k$, followed by a subsampling operation with stride $s$. \textit{Middle}: Inclusion of a low-pass filter before subsampling. \textit{Bottom}: Adaptive Polyphase Sampling (APS) can be used instead of naive subsampling.}
  \label{fig:diagram}
\end{figure}

Our focus is on evaluating mechanisms to improve shift invariance applied to the subsampling operations within max-pooling layers in CNNs.
A max-pooling layer with squared size $k$ and stride $s$ can be understood as the cascade of two operations, as illustrated in the top diagram of Fig.~\ref{fig:diagram}: a densely-evaluated (i.e., with unit stride) max-pooling operation with size $k$, followed by a subsampling operation with stride $s$ greater than unity.

\subsection{Low-Pass Filtering Before Subsampling}
\label{sec:lpf}
We focus on the effect of low-pass filtering feature maps before subsampling in the context of a max-pooling layer, inspired by Zhang~\cite{zhang2019making}.
The subsampling operation may incur in aliasing problems as the incoming signal (the feature map) is not band-limited.  
The classic signal-processing fix is to add a low-pass filter before subsampling \cite{oppenheim2001discrete}.
One manner to realize this filter is through a 2D kernel, $LPF_{m,n}$, of size $m$ x $n$, such that the max-pooling layer for an incoming feature map $x$ becomes
\begin{equation}
  \label{eqn:blurpool}
        y_{lpf} = Subsample_s(LPF_{m,n}(MaxPool_{k,1}(x))),
\end{equation}
where $MaxPool_{k,1}$ is a max-pooling operation across areas of size $k$ x $k$ and unit stride, $LPF_{m,n}$ applies a low-pass filter of size $m$ x $n$, and $Subsample_s$ denotes naive subsampling with a stride $s$, as illustrated in the middle diagram of Fig.~\ref{fig:diagram}.

This simple measure can have different benefits when applied within CNNs.
First, in case the feature maps present energy variations of too high frequency for the subsampling operation to be carried errorless, $LPF_{m,n}$ will help mitigate aliasing.\footnote{By high-frequency energy variations in the feature map we refer to rapid spectro-temporal modulations or sharp patterns in the 2D signal formed by a feature map.
This should not be confused with the frequency components of the input audio signal.
The high-frequency energy variations are not necessarily constrained to a specific region of the feature map.
For example, a sequence of human clapping sounds 
forms a time-frequency representation with a series of transients. 
In its corresponding feature map, the energy variations given by such a sequence of transients can generate high frequencies, even at the lowest end of the spectrum.}
This can reduce the amount of corrupted information flowing through the network.
Second, the signal processing literature demonstrates how preventing aliasing can favour shift invariance in a given process \cite{oppenheim2001discrete}. 
One way to see it is that $LPF_{m,n}$ spreads possible sharp patterns across neighbouring feature map bins.
Intuitively, when subsampling differently shifted versions of a spectrogram, the subsampled feature maps are likely to be more structurally similar if they have been previously low-pass filtered.
This could provide the network with improved generalization to this kind of small shifts, potentially increasing classification performance.
Third, $LPF_{m,n}$ is essentially blurring or smoothing out the incoming feature map, which could be understood as a form of regularization.
For example, L2 regularization is a common way to penalize outlier weights with large absolute values, driving them close towards zero \cite{cortes2012l2}.
It could be argued that the proposed $LPF_{m,n}$ inflicts a similar effect on the feature map bins, smoothing out the most drastic energy variations---in other words, attenuating the high frequency components in the 2D signal formed by the feature map.
In Sec. \ref{sec:experiments} we discuss through experiments which of these hypotheses seem more plausible.

To implement $LPF_{m,n}$, some of the most basic characteristics of common 2D image-oriented low-pass filter kernels are: non-negative weights that add up to unity \cite{distante2020handbook}.
This can be realized in several ways.

\smallskip
\noindent \textbf{Non-trainable low-pass filters.}
These are commonly defined as binomial filters, which are in turn discrete approximations of Gaussian filters.
To generate 1D binomial filters, a simple manner is to repeatedly convolve the base averaging mask [1,1] with itself, in order to get filter masks such as [1,2,1], [1,3,3,1], or [1,4,6,4,1], for one, two and three convolutions, respectively.
Then, a 2D squared binomial mask can be obtained simply by convolving a 1D binomial filter with its transpose \cite{distante2020handbook}.
In Sec. \ref{sec:experiments} we denote this type of filters as \textit{BlurPool} for consistency with \cite{zhang2019making} as the non-trainable low-pass filters that we use in our experiments are largely inspired by this work.

\smallskip
\noindent \textbf{Trainable low-pass filters.}
These can be defined by randomly initializing a kernel with dimensions $m$ x $n$, and learning their weights through back propagation. 
In order to imprint the low-pass nature to the filter, its weights can be passed through a softmax function to ensure non-negativity and normalization. 
In Sec. \ref{sec:experiments} we denote this type of filters as Trainable Low-Pass Filter (\textit{TLPF}).
An alternative is to create auxiliary loss functions to encourage the filter weights to adopt a low-pass behaviour through loss penalization.
Trainable low-pass filters have also been used recently within a learnable audio frontend \cite{zeghidour2021leaf}.

In this work, we compare non-trainable low-pass filters (BlurPool) and trainable low-pass filters (TLPF) constrained via softmax, for simplicity.
A given low-pass filter can be applied over the incoming feature map via a convolution operation that also incorporates the required subsequent subsampling stride $s$.
More specifically, $Subsample_s(LPF_{m,n}(\cdot))$  in (\ref{eqn:blurpool}) is implemented via a depthwise separable convolution using trainable or non-trainable $LPF_{m,n}$ and stride $s$.


\subsection{Adaptive Polyphase Sampling}
\label{sec:aps}
Adaptive polyphase sampling (APS) is a downsampling mechanism that directly addresses the lack of shift invariance caused by subsampling operations \cite{chaman2021truly}.
The underlying principle of APS is based on a simple observation: the result of subsampling a time-frequency (T-F) patch and subsampling its shifted-by-one-bin version can be different when bins are sampled at the same fixed positions.
This happens because the energies captured by the same grid over two shifted patches are likely to be different.
However, when subsampling a feature map, multiple candidate grids could actually be used instead of always using the same grid (as typically done).
Intuitively, a time/frequency shift applied over an input patch could be seen conceptually as translating its energy bins from one grid to another.
One way to be robust to these shifts is to select the subsampling grid adaptively based on some criterion, such that the grid follows the shift at the input.

More formally, given an input feature map $x$, and considering a subsampling operation\footnote{This subsampling operation would follow a densely-evaluated max pooling operation in order to form a typical max-pooling layer, see Fig. \ref{fig:diagram}} with stride 2, there are four possible grids that can be used for subsampling, depending on which bin from the four options in each 2x2 area is passed to the output.
Subsampling with each grid will yield one of the four possible candidate subsampled feature maps, termed \textit{polyphase components} \cite{chaman2021truly}, which can be denoted as $\left\{y_{ij}\right\}_{i,j=0}^1$.
Analogously, if we consider a shifted-by-one-bin version of the input feature map, $\tilde{x}$, its polyphase components are given by $\left\{\tilde{y}_{ij}\right\}_{i,j=0}^1$.

The conventional course of action consists of always choosing the same subsampling grid and consequently returning the same polyphase component (e.g., $y_{00}$ by picking the top left bin in each 2x2 area).
However, as mentioned, depending on the input patch, this will likely cause different downsampled outputs when the patch is simply shifted by one bin ($y_{00} \neq \tilde{y}_{00}$).
It can be demonstrated that the set $\left\{\tilde{y}_{ij}\right\}$ is a re-ordered version of $\left\{y_{ij}\right\}$ \cite{chaman2021truly} (which may be potentially shifted, but carrying identical energy values).
Therefore, by adaptively choosing a polyphase component in a permutation invariant way, a very similar subsampled output, $y_{i_{aps}j_{aps}}$, would be obtained regardless of sampling from $x$ or $\tilde{x}$.
The adaptive selection can be done by maximizing a given criterion, for example, maximizing some norm $l_p$, as given by
\begin{equation}
  \label{eqn:aps}
        i_{aps},j_{aps} = \argmax_{i,j} \left\{\norm{y_{ij}}_p\right\}_{i,j=0}^1,
\end{equation}
where $p\in\left\{1,2\right\}$.
In this way, by substituting the naive subsampling in a max-pooling layer by APS (as illustrated in the bottom diagram of Fig.~\ref{fig:diagram}), robustness to incoming time/frequency shifts is increased.

The benefit from APS comes from the generalization to shifts embedded in the network’s architecture, in a fashion conceptually similar to what is done with $LPF_{m,n}$ (Sec. \ref{sec:lpf}).
APS, in contrast, provides no explicit measures against potential aliasing problems.



\section{Experimental Setup}
\label{sec:setup}

\subsection{Evaluation and Training Details}
\label{ssec:dataset}
We evaluate the proposed methods on a large-vocabulary sound event tagging task using the recently released FSD50K dataset \cite{fonseca2020fsd50k}.
FSD50K is open dataset of sound events containing over 51k Freesound\footnote{\url{https://freesound.org}} audio clips, totalling over 100h of audio manually labeled using 200 classes drawn from the AudioSet Ontology \cite{gemmeke2017audio}.
We follow the evaluation procedure proposed in the FSD50K paper \cite{fonseca2020fsd50k} (with minor deviations), which is outlined next.
Incoming clips are transformed to log-mel spectrograms using a 30ms Hann window with 10ms hop, and 96 bands.
To deal with the variable-length clips, we use T-F patches of 1s, equivalent to 101 frames, yielding patches of $T \times F=101 \times 96$ that feed the networks. Clips shorter than 1s are replicated while longer clips are trimmed in several patches with 50\% overlap inheriting the clip-level label.
We train, validate and evaluate using the proposed \textit{train} set, \textit{val} set and \textit{eval} set \cite{fonseca2020fsd50k}.
Models are trained using Adam optimizer \cite{kingma2014adam} to minimize binary cross-entropy loss.
Learning rate is 3e-5, halved whenever the validation metric plateaus for 10 epochs.
Models are trained up to 150 epochs, earlystopping the training whenever the validation metric is not improved in 20 epochs. We use a batch size of 128 and shuffle training examples between epochs. Once the training is over, the model checkpoint with the best validation metric is selected to predict scores and evaluate performance on the eval set.
For inference, we compute output scores for every (eval or val) T-F patch, then average per-class scores across all patches in a clip to obtain clip-level predictions.
Our evaluation metric is balanced mean Average Precision (mAP), that is, AP computed on a per-class basis, then averaged with equal weight across all classes to yield the overall performance, following \cite{fonseca2020fsd50k,gemmeke2017audio,fonseca2020addressing}.

\subsection{Baseline Model}
\label{ssec:baseline}
As a base network, we use a VGG-like architecture \cite{simonyan2014very}.
This type of architecture has been widely used for SET \cite{kong2019panns,dorfer2018training,hershey2017cnn,ebrahimpour2020end} and is the most competitive baseline for FSD50K when compared to others of higher complexity \cite{fonseca2020fsd50k}---which also accords with recent music tagging evaluations \cite{won2020evaluation}.
Due to its limited size compared to other baselines \cite{fonseca2020fsd50k}, it allows faster experimentation.
In addition, this architecture conveniently features several max-pooling layers that allow the study of the proposed pooling mechanisms.
Specifically, the network that we use for the majority of our experiments is similar to a VGG type A \cite{simonyan2014very}. 
The network, denoted as \textit{VGG41}, consists of 4 convolutional blocks, with each block comprising two convolutions with a receptive field of (3,3), and each convolution followed by Batch Normalization \cite{ioffe2015batch} and ReLU activation. 
Between the blocks, max-pooling layers of size (2,2) (and same stride) are placed by default---they will be substituted by the proposed pooling mechanisms.
A densely-evaluated max pooling operation (of size 3x3 and unit stride) will sometimes be inserted between the convolutions within each block---we will refer to it as \textit{intra-block pooling} (IBP).
This provides partial translation invariance but not dimensionality reduction, allowing the same (max) element to be transferred to the output in adjacent spatial locations.
This tweak has been applied in various non-audio applications \cite{goodfellow2016deep}, and to a lesser extent also in SET tasks \cite{ebrahimpour2020end}.
Finally, in order to summarize the final feature map information before the output classifier, we use a global pooling in which we first aggregate information along the spectral dimension via averaging for every time step, then max-pool the outcome in the time dimension.
We found out that aggregating first spectral and then temporal information in this manner is the most beneficial for our task among other combinations.
VGG41 has 1.2M weights, which allows for relatively fast experimentation. 
The baseline and topline settings using VGG41 are also evaluated using \textit{VGG42} (of 4.9M weights), where we double the width of the network with respect to VGG41 (i.e., using twice the number of filters in every convolutional layer).

\subsection{\textit{mixup}}
\label{ssec:mixup}
We evaluate the baseline and top performing methods proposed in Sec. \ref{sec:method} with or without \textit{mixup} augmentation \cite{zhang2017mixup} in order to analyze their behavior in presence of a strong regularizer.
Mixup acts as a regularizer by encouraging networks to predict less confidently on linear interpolations of training examples.
In particular, it augments the training distribution by creating virtual examples under the assumption that linear interpolations in the feature space correspond to linear interpolations in the label space.
Following \cite{zhang2017mixup}, we sample $\lambda$ from a beta distribution $\lambda \sim$ Beta$(\alpha, \alpha)$, for $\alpha \in (0, \infty)$.
The hyper-parameter $\alpha$, which controls the interpolation strength, is set to 1.25 after tuning on the val set.

We choose mixup because the concept of mixing sounds is an audio-informed operation, and it has been proven useful for SET \cite{kong2019panns,Fonseca2019model,gong2021psla} and other sound event research tasks \cite{fonseca2021unsupervised}. 
In our view, mixup can be interpreted from two different perspectives. 
First, it is a regularizer to mitigate overfitting, which can be important at our scale of data, especially for some classes that present less than 100 training clips.
Second, mixup is a mechanism that allows to cover during training a diversity of examples that may be encountered in evaluation, hence improving generalization.
In particular, upon the creation of FSD50K, audio clips with multiple sound sources were prioritized to some extent for the eval set, whereas the dev set presents a higher proportion of single-source clips.
It can therefore be argued that a kind of domain shift exists between both sets, which is being partially compensated through mixup.
Hence, this type of augmentation is specially well aligned with the recognition task of FSD50K.

\section{Experiments}
\label{sec:experiments}
We evaluate the methods proposed in Sec. \ref{sec:method} on the SET task posed by FSD50K, using VGG41 (Sec. \ref{ssec:eval_41}) and also using mixup and VGG42 (Sec. \ref{ssec:eval_42}).
In Sec. \ref{ssec:quanti} we demonstrate that the methods are increasing the network's robustness to input shifts.
Sections \ref{ssec:discuss} and \ref{ssec:exp_previous_work} provide discussion and comparison with previous work on FSD50K.
For all the performance results (Tables \ref{tab:results} and \ref{tab:progress}), we report average performance and standard deviation across three runs.

\subsection{Evaluation using a Small Model}
\label{ssec:eval_41}
\begin{table*}[ht]
\caption{mAP obtained by inserting different pooling mechanisms into the VGG41 baseline. TLPF = Trainable Low-Pass Filter, APS = Adaptive Polyphase Sampling, IBP = Intra-block Pooling.}
\vspace{-1mm}
\centering
\begin{tabular}{lc|lc|lc}
\toprule
\textbf{Method} & \textbf{mAP} & \textbf{Method}   & \textbf{mAP}  & \textbf{Method}  & \textbf{mAP}  \\
\midrule
\midrule
VGG41 (baseline)           & 0.457 $\rpm$ 0.003   & + BlurPool 5x5 + IBP   & 0.479 $\rpm$ 0.003  & + TLPF 3x3 + IBP    & 0.478  $\rpm$ 0.002                               \\
+ BlurPool 3x3             & 0.475 $\rpm$ 0.002 	& + TLPF 5x5 + IBP    & 0.481  $\rpm$ 0.002  & + TLPF 4x4 + IBP    & 0.480  $\rpm$ 0.004    \\
+ BlurPool 5x5             & 0.476 $\rpm$ 0.003	    & + TLPF 5x5 + APS $l_1$ & \textbf{0.484} $\rpm$ 0.002 & + TLPF 5x5 + IBP    & \textbf{0.481}  $\rpm$ 0.002      \\
+ TLPF 3x3                & 0.476 $\rpm$ 0.003	    & + APS $l_1$ + IBP   & 0.478 $\rpm$ 0.001  & + TLPF 6x6 + IBP    & 0.480  $\rpm$ 0.001\\
+ TLPF 5x5                & 0.479 $\rpm$ 0.003      & &  & + TLPF 1x4 + IBP    & 0.475  $\rpm$ 0.005                   \\
+ TLPF 6x6                & 0.477 $\rpm$ 0.001      & &  & + TLPF 1x5 + IBP    & 0.480  $\rpm$ 0.001                   \\
+ APS $l_1$              & \textbf{0.480} $\rpm$ 0.001	   & & & + TLPF 1x6 + IBP    & 0.480  $\rpm$ 0.002                  \\
+ APS $l_2$               & 0.460  $\rpm$ 0.002    & & & + TLPF 4x1 + IBP    & 0.469  $\rpm$ 0.004                            \\
+ IBP                     & 0.472 $\rpm$ 0.002 	   & & & + TLPF 5x1 + IBP    & 0.470  $\rpm$ 0.004                             \\
                         & 	                       & & & + TLPF 6x1 + IBP    & 0.472  $\rpm$ 0.002                             \\
\bottomrule
\end{tabular}
\label{tab:results}
\end{table*}

Table \ref{tab:results} shows the results of inserting the pooling mechanisms individually into VGG41 (left section) as well as in some pairwise combinations (center section).
The right section lists the results of exploring TLPF in depth.
By looking at the left section, it can be seen that all the evaluated methods outperform the baseline system.
That is, inserting each of the methods alone into a standard VGG-like architecture improves recognition performance.
The boosts range from 0.003 in the worst case (APS $l_2$) to 0.023 in the best case (APS $l_1$).
If we focus on the low-pass filter based solutions, we observe that this classical signal processing technique is beneficial for CNN-based sound event classification.
While it may seem that blurring the feature maps can smooth out relevant detailed information (thus leading to performance degradation) results indicate that it is indeed helpful.
The choice of trainable vs non-trainable low-pass filters does not seem critical, yet the trainable version TLPF seems to produce slightly higher mAP values.
The different sizes of these filters allow to find a trade-off between high-frequency smoothing and loss of information in the incoming feature maps (the larger the size, the stronger the smoothing effect).
Results seem to indicate that larger smoothing areas (5x5 vs 3x3) are beneficial.
By looking at results with APS, we observe that $l_1$ outperforms $l_2$ as norm criterion.
We also did preliminary experiments with other metrics such as $l_\infty$, $l_0$ and variance, but we found $l_1$ to be the best choice overall.
Interestingly, a naive tweak like IPB also shows some impact, although more modest than that of the other methods.
The two top methods when applied individually are APS $l_1$ and TLPF 5x5, showing on par performance.

We set out to combine some of the methods in pairs in order to see if they are complementary (center section).
Combining low-pass filtering (which operates before subsampling between convolutional blocks) and IBP (which operates between convolutions within every block) seems to provide a small but consistent boost, for both BlurPool and TLPF.
When joining the top performing methods, specifically, low-pass filtering the incoming feature maps with TLPF 5x5, followed by subsampling them with APS, we observe a small performance boost.
A possible explanation for their complementarity could lie in TLPF addressing aliasing issues while APS is agnostic to it.
Finally, joining APS and IBP does not yield further boosts.

The right section of Table \ref{tab:results} shows the results of exploring different low-pass filter shapes in TLPF.
In previous work, low-pass filters are usually adopted for computer vision tasks, hence they are of squared size (e.g., \cite{zhang2019making,vasconcelos2020effective}).
Here, we seek to find out if there is one axis of the audio spectrogram feature maps (time or frequency) along which low-pass filtering is more beneficial.
To this end, we run experiments using 1D trainable low-pass filters applied only along the frequency axis (filters of size $1$ x $n$) or the temporal axis (filters of size $m$ x $1$).
At the top of the right section, we first report the results by progressively increasing the area of squared filters.
A sweet spot in the size 5x5 can be observed.
Then we report results by low-pass filtering only along the frequency axis.
Interestingly, we find out that much of the performance obtained with squared filters is already achieved by smoothing out the spectral variations alone.
In contrast, when we apply the low-pass filters only along the time axis the performance is noticeably worse.

\subsection{Evaluation using Regularization and a Larger Model}
\label{ssec:eval_42}
Next, we select the best setups of the two pooling mechanisms considered on VGG41 (one based on low-pass filtering and another based on APS), as well as their combination.
Table \ref{tab:progress} shows the results using these setups, now adding mixup augmentation and also doubling the width of the network, which means multiplying its number of weights approximately by four.
\begin{table}[!t]
\vspace{-2mm}
\caption{mAP obtained by using top performing pooling mechanisms in presence of mixup and with the larger capacity VGG42. Values in parenthesis are absolute improvements over the corresponding baseline. TLPF = Trainable Low-Pass Filter, APS = Adaptive Polyphase Sampling, IBP = Intra-block Pooling.}
\vspace{-1mm}
\centering
\begin{tabular}{@{}lccc@{}}
\toprule
                & \textbf{VGG41} & \textbf{VGG41} & \textbf{VGG42}  \\
\textbf{Method} &                & \textbf{+ mixup}  & \textbf{+ mixup}   \\
\midrule
\midrule
Baseline                 & 0.457 $\rpm$ 0.003       & 0.497 $\rpm$ 0.003          & 0.523 $\rpm$ 0.002\\
\midrule
+ APS $l_1$              & 0.480 $\rpm$ 0.001	    & 0.513  $\rpm$ 0.003          & 0.538 $\rpm$ 0.004 \\
                         & \footnotesize{(+0.023)}	& \footnotesize{(+0.016)}      & \footnotesize{(+0.015)} \\
+ TLPF 5x5 + IBP         & 0.481  $\rpm$ 0.003     & 0.511  $\rpm$ 0.003           & 0.539 $\rpm$ 0.002\\
                         & \footnotesize{(+0.024)} & \footnotesize{(+0.014)}       & \footnotesize{(+0.016)} \\
+ TLPF 5x5 + APS $l_1$   & \textbf{0.484} $\rpm$ 0.002      & \textbf{0.514} $\rpm$ 0.003            & \textbf{0.541}  $\rpm$ 0.002   \\
                         & \footnotesize{(+0.027)} & \footnotesize{(+0.017)}       & \footnotesize{(+0.018)} \\
\bottomrule
\end{tabular}
\label{tab:progress}
\vspace{-4mm}
\end{table}
The left column lists the best values from Table \ref{tab:results}, showing boosts from 0.023 to 0.027 with respect to the baseline.
When we add mixup to VGG41 (center column) substantial performance improvements are observed, demonstrating the good alignment of this operation with SET in general (which accords with \cite{kong2019panns,gong2021psla}) and with the FSD50K classification task in particular (as discussed in Sec. \ref{ssec:mixup}).
All methods perform in the same ballpark, showing boosts with respect to the baseline of up to 0.017, with the combination of TLPF and APS yielding top mAP.

Our motivation to combine the proposed methods with mixup is to analyze their behaviour in presence of a strong regularizer.
If they act solely as a general form of regularization, we would expect them to provide limited boosts when combined with mixup.
We do observe a certain improvement decrease when combined, but the boosts are still solid, both when using VGG41 and VGG42 (center and right columns).
These results suggest that the proposed methods are addressing problems beyond lack of regularization, presumably reinforcing robustness to time/frequency shifts at the input.
In Sec. \ref{ssec:quanti} we demonstrate that this is the case by systematically applying time/frequency shifts to a set of input spectrogram patches, and analyzing the network's robustness against these shifts with and without the proposed pooling mechanisms.
Finally, when inserting these pooling methods into VGG42 in presence of mixup (right column), we see that they are also beneficial within a larger-capacity model where performance is more competitive (in our case, increasing the capacity from 1.2 to 4.9M weights).
In particular, combining TLPF and APS yields the top mAP again, showing a boost of 0.018 over the baseline.

\subsection{Characterizing the Increase of Shift Invariance}
\label{ssec:quanti}
In previous experiments we have seen that classification performance is improved when we adopt the proposed pooling mechanisms, presumably due to the increase of shift invariance.
Here, we demonstrate empirically that these pooling mechanisms are indeed addressing this problem.
To this end, we apply shifts to a set of input spectrogram patches and analyze the network's robustness against these shifts with and without the proposed pooling mechanisms.
The set of data for this evaluation consists of 1000 audio clips\footnote{This list of 1000 files is released for future evaluations.} from the eval set.
They are selected at random after applying the following constraints for a more controlled experimental scenario: \textit{i)} we choose clips with one single label (i.e., presumably containing one single sound event); \textit{ii)} we choose clips with a minimum length of 2s, so that we can always select a patch in the time span [0.5, 1.5] s, allowing the discard of potential preceding silence that sometimes occurs.
The shifts applied to the input 1s T-F patches of $T \times F=101 \times 96$ obey one of the two following protocols.
The first protocol (denoted as \textit{time-$n_f$}) is simply a time shift of the patch by $n_f$ frames, with $n_f \in \left\lbrace 1,3,5\right\rbrace$. 
Each unity of $n_f$ corresponds to 10ms (the hop size when framing the input audio signal).
The second protocol (denoted as \textit{freq-$n_b$}) consists of shifting the input patch by $n_b$ mel bands upwards in frequency, with $n_b \in \left\lbrace 1,3,5\right\rbrace$.
The $n_b$ original highest bands are discarded, and the $n_b$ lowest bands in the new patch are filled with white noise centered at the mean value of the original lowest band. 
By doing this, we analyze the effect not only of frequency shifts but also of small artificial perturbations in the input.

For every input patch, we \textit{i)} apply one shift according to one of the protocols above; \textit{ii)} compute network predictions for both original and shifted patches, and \textit{iii)} measure the predictions’ sensitivity to the shift using two metrics, namely, \textit{classification consistency} and \textit{mean absolute change}.
Classification consistency refers to the percentage of cases in which the network predicts the same top class for both original and shifted patches \cite{zhang2019making,chaman2021truly}.
Mean absolute change (MAC) is a metric that measures the absolute change of the probability predicted for the top class after the shift, averaged across the 1000 examples \cite{azulay2018deep}. 
The motivation to use MAC is to rule out the possibility that variations in classification consistency are originated by tiny differences between the top class and the second most likely class predicted.

Table \ref{tab:quantify} shows the result of this evaluation for the time and frequency protocols.
In Table \ref{tab:quantify}, \textit{baseline} corresponds to the VGG41 baseline of Table \ref{tab:results}, whereas \textit{proposed} corresponds to the same model after incorporating TLPF 5x5 and APS $l_1$.
\begin{table}[!t]
\caption{Classification consistency (in \%, higher is better) and mean absolute change (MAC) (lower is better) when applying time and frequency shift protocols over input patches. Models evaluated are the \textit{baseline} of Table \ref{tab:results} and the same model after inserting TLPF 5x5 and APS $l_1$ (\textit{proposed}). The proposed model exhibits higher robustness to shifts.}
\vspace{-1mm}
\centering
\begin{tabular}{lcc|cc}
\toprule
\textbf{Protocol} & \multicolumn{2}{c}{\textbf{Consistency (\%)}}   & \multicolumn{2}{c}{\textbf{MAC}} \\
                       	    & baseline & proposed       & baseline & proposed \\
\midrule
\midrule
time-1       	    & 82.0 & 92.5       & 0.078 & 0.030 \\
time-3              & 75.2 & 87.2       & 0.106 & 0.048 \\
time-5              & 74.0 & 83.2       & 0.110 & 0.062 \\

\midrule
freq-1        	    & 63.5  & 79.8         & 0.175 & 0.078 \\
freq-3        	    & 53.3 & 67.6         & 0.234 & 0.145 \\
freq-5        	    & 49.0 & 60.0         & 0.263 & 0.199 \\

\bottomrule
\end{tabular}
\label{tab:quantify}
\vspace{-4mm}
\end{table}
In the results of the time protocol (top section), we see that by applying a time shift of only 1 frame (10 ms), the baseline network changes its top prediction 18\% of the time.
Note that this is a minor modification in the time framing of the input audio signal, which can be regarded as imperceptible to a human in most cases.
The proposed network (including the mechanisms to increase shift invariance) shows higher classification consistency than the baseline for all the cases considered.
As the time shifts increase, the network becomes less consistent (according to our definition).
These findings are confirmed by the proposed network consistently showing smaller MAC values, i.e., the output probability for the top class is more robust to the time shifts.
By looking at the results of the frequency protocol (bottom section) we observe a similar trend, with the proposed network showing increased robustness to frequency shifts and small perturbations (larger classification consistency percentages, and smaller MAC values than the baseline model).
Interestingly, we see that the classification consistency values in this protocol are overall lower than those observed for the time protocol (and vice versa for the MAC values).
For example, the minimal shift applied in time yields consistency values of 82.0\% and 92.5\% for baseline and proposed models, respectively.
In contrast, the minimal shift applied in frequency leads to analogous values of 63.5\% and 79.8\%.
This can be due to several reasons.
First, the network considered may be more sensitive to frequency shifts than to time shifts. 
This could be linked to results in Sec. \ref{ssec:eval_41}, where low-pass filtering along the frequency axis is shown to be more effective than along time.
Second, it could happen that the frequency shifts affect the semantics of the input examples to some degree (which is more unlikely with the time protocol).
Third, in this protocol we are introducing small artificial perturbations never seen at training time, which may confuse the network.
Regardless, the proposed approach exhibits higher robustness to the applied shifts in all cases analyzed.
Similar trends are observed when using the proposed methods alone (either TLPF or APS).
In summary, results of Table \ref{tab:quantify} demonstrate that the proposed pooling mechanisms increase shift invariance in the network.

\begin{figure}[t]
  \centering
  \centerline{\includegraphics[width=1.05\columnwidth]{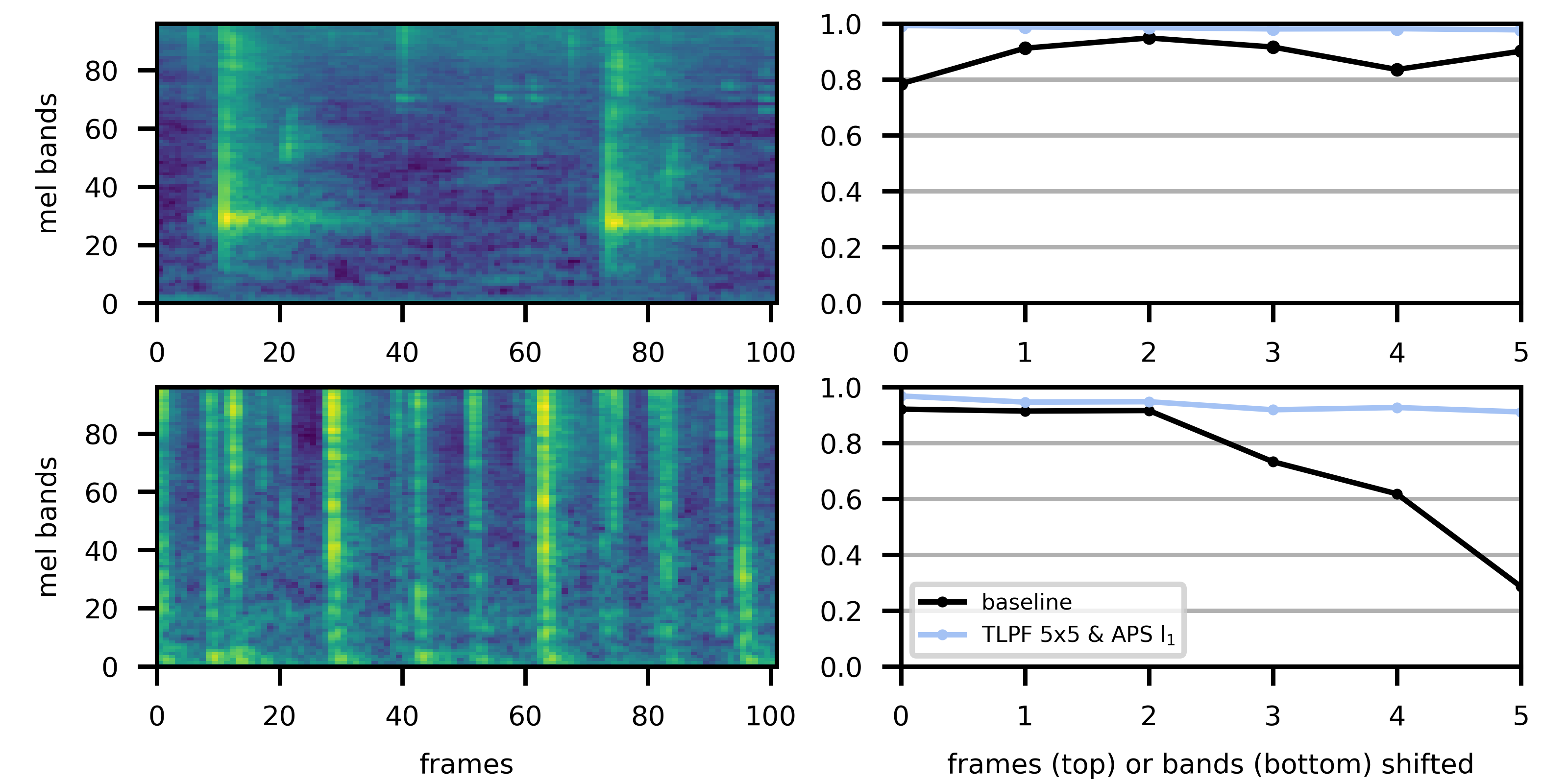}}
    \caption{Predicted score for the correct class of water dripping (top) and computer keyboard typing (bottom) examples, as a function of shifted time frames (top) and mel bands (bottom). Inserting the pooling mechanisms (TLPF 5x5 + APS $l_1$) makes the predictions more stable against spectrogram shifts.}
  \label{fig:shift_ex}
\end{figure}

Fig. \ref{fig:shift_ex} shows the classification stability of the same models used for Table \ref{tab:quantify} with two examples: applying the time shift protocol over a water dripping sound (top) and the frequency shift protocol over a computer keyboard typing sound (bottom).
In the top plots, predictions for the \textit{Drip} class are stable when we insert the proposed pooling mechanisms, as one would expect upon shifting the signal framing by few miliseconds. 
However, the baseline predictions show certain fluctuations.
Similarly, in the bottom plots the proposed network exhibits higher robustness against the frequency shifts and the induced perturbations.



\subsection{Discussion}
\label{ssec:discuss}
We have seen that two methods with different underlying principles targeting the increase of shift invariance yield improvements within the same ballpark in our task.
Further, we have empirically shown that they increase model's robustness to spectrogram shifts.
These facts demonstrate that there is indeed some lack of this property in the CNN under test, and suggest that reinforcing shift invariance is beneficial for sound event classification.
One interesting observation is that while anti-aliasing measures are helpful to increase performance and shift invariance, they do not seem strictly necessary in light of the overall similar performance attained by APS.

In terms of model size, the impact is negligible for all evaluated methods.
Specifically, TLPF5x5 adds 6k (0.50\%) and 12k (0.24\%) trainable parameters over VGG41 and VGG42 respectively.
Its non-trainable counterpart (BlurPool 5x5) adds the same number of non-trainable parameters.
APS does not require any additional parameters (trainable or non-trainable).
The additional compute required by the methods is also limited.
For the low-pass filtering methods, one additional convolution is needed to apply the low-pass filter over the incoming feature maps for every subsampling operation.
Analogously, the only additional compute required by APS is the computation of the polyphase components and their norms in every subsampling operation.
Thus, the proposed architectural modifications (which apply only to the pooling layers) yield consistent recognition boosts when inserted into a well-known CNN, with minimal additional computation.
This makes them an appealing alternative to conventional pooling layers.

\subsection{Comparison with Previous Work}
\label{ssec:exp_previous_work}
Table \ref{tab:sota} lists previously published results on FSD50K, including mAP and d-prime ($d'$) when available.
Our best system obtains state-of-the-art mAP of 0.541, slightly outperforming recent Transformer-based approaches (0.537) \cite{verma2021audio}, as well as the PSLA approach when trained only on FSD50K (0.452) \cite{gong2021psla}.
PSLA makes use of a collection of training techniques (ImageNet pretraining, data balancing and augmentation, label enhancement, weight averaging, and ensemble of several models) \cite{gong2021psla}.
Among all of them, the key ingredient seems to be ImageNet pretraining, without which the performance decreases dramatically.
While using transfer learning from ImageNet seems to provide substantial boosts, we consider transfer learning from external datasets a different track.
Our proposed state-of-the-art approach consists of simple architectural changes inserted into a widely-used CNN at minimal computational cost along with simple augmentation.
\begin{table}[!t]
\vspace{-2mm}
\caption{State-of-the-art on FSD50K.}
\vspace{-1mm}
\centering
\begin{tabular}{lcc}
\toprule
\textbf{Method}                             & \textbf{mAP}  & \pmb{$d’$} \\
\midrule
\midrule
Baseline \cite{fonseca2020fsd50k}                        & 0.434      & 2.167   \\
\midrule
PSLA (not using ImageNet)  \cite{gong2021psla}          & 0.452      & - \\
Audio Transformers \cite{verma2021audio}                & 0.537        & -	        \\
VGG42 + APS $l_1$ (ours)                                & 0.538        & 2.415     \\
VGG42 + TLPF5x5 + IBP (ours)                            & 0.539        & 2.417    \\
VGG42 + TLPF 5x5 + APS $l_1$  (ours)                    & \textbf{0.541} &\textbf{2.431} \\
\midrule
\textcolor{mygray}{PSLA (using ImageNet)} \cite{gong2021psla} & \textcolor{mygray}{0.567} & - \\
\bottomrule
\end{tabular}
\label{tab:sota}
\vspace{-4mm}
\end{table}

\section{Conclusion}
\label{sec:conclu}
We have evaluated two pooling methods to improve shift invariance in CNNs in the context of a sound event classification task.
These methods are based on low-pass filtering and adaptive sampling of incoming feature maps, and are implemented via small modifications in the pooling layers of CNNs.
We have evaluated the effect of these architectural changes on the FSD50K dataset, using models of different capacity and in presence of strong regularization.
Results show that the models evaluated indeed present a problem of only-partial shift invariance, and that adopting the proposed methods to improve it yields recognition boosts.
The improvements observed are within the same ballpark for both methods, despite them having different underlying principles, which allows for small further boosts via their combination.
Inserting these pooling methods into VGG variants makes the networks exhibit higher robustness to time/frequency shifts in the input spectrograms.
These facts suggest that reinforcing shift invariance in CNNs is beneficial for sound event classification.
The proposed architectural changes applied to a widely-used CNN yield consistent recognition improvements with minimal additional computation, which makes them an appealing alternative to conventional pooling layers.
Our best system achieves a new state-of-the-art mAP of 0.541 on FSD50K.





\newpage
\bibliographystyle{IEEEtran}
\bibliography{refs}

\end{document}